\documentclass[preprint,showpacs,twoside,preprintnumbers,nofootinbib]{revtex4}
\usepackage{graphicx,amsmath,amsfonts,amssymb,amsbsy}
\usepackage{pstricks}
\usepackage{mathrsfs}
\usepackage{pdfpages}
\usepackage{slashed}

\begin{document}

\title{Phase diagrams in nonlocal Polyakov--Nambu--Jona-Lasinio models constrained by Lattice QCD results}

\author{G. A. Contrera$^{1,2,3}$, A. G. Grunfeld$^{2,4}$,
D. B. Blaschke $^{3,5}$}

\affiliation{$^1$ Gravitation, Astrophysics and Cosmology Group, FCAyG, UNLP,
La Plata, Argentina\\
   $^2$ CONICET, Rivadavia 1917, 1033 Buenos Aires, Argentina\\
   $^3$ Institute for Theoretical Physics, University of Wroc{\l}aw,
50-204 Wroc{\l}aw, Poland\\
   $^4$ Departamento de F\'\i sica, Comisi\'on Nacional de Energ\'{\i}a
At\'omica, (1429) Buenos Aires, Argentina\\
   $^5$ Bogoliubov Laboratory for Theoretical Physics, JINR Dubna,
141980 Dubna, Russia}

\begin{abstract}
Based on lattice QCD-adjusted $SU(2)_f$ nonlocal Polyakov--Nambu--Jona-Lasinio
(PNJL) models, we investigate how the location of the critical endpoint in the
QCD phase diagram depends on the strenght of the vector meson coupling, as well
as the Polyakov-loop (PL) potential and the form factors of the covariant
model.
The latter are constrained by lattice QCD data for the quark propagator.
The strength of the vector coupling is adjusted such as to reproduce the slope
of the pseudocritical temperature for the chiral phase transition at low
chemical potential extracted recently from lattice QCD simulations.
Our study supports the existence of a critical endpoint in the QCD phase
diagram albeit the constraint for the vector coupling shifts its location to
lower temperatures and higher baryochemical potentials than in
the case without it.
\end{abstract}

\date{\today}
\maketitle

\section{Introduction}
%

The QCD phase diagram has been focus of intense research in the last decades.
The conjecture for the existence of a critical endpoint (CEP) of first order
phase transitions in the QCD phase diagram is the basis for recent as well as
future beam energy scan (BES) programs in relativistic heavy-ion collision
experiments at RHIC, SPS, NICA and fair which try to identify the parameters
of its position $(T_{\rm CEP}, \mu_{\rm CEP})$.
The theoretical situation is very unsatisfactory since the predictions for
this position form merely a skymap in the $T-\mu$ plane
\cite{Stephanov:2007fk}.
Quantitative calculations of the phase diagram based in QCD are extremely
complicated in the low energy regime due to its strong coupling.
In this region non-perturbative methods are powerful tools to describe the
chiral and deconfinement transitions.
Lattice QCD calculations have the sign problem at finite chemical potential.
Therefore, the effective models play a crucial role to describe the
phase diagram, specially at finite densities.

The now well-established results from lattice QCD at zero and small
chemical potential $\mu$, predict coincident chiral and deconfinement
crossover transitions
at a pseudocritical temperature of  $T_c(0)=154 \pm 9$ MeV for 2+1 flavors
\cite{Bazavov:2011nk} and a value of $T_c(0)\sim 170$ MeV for two flavors
\cite{Ejiri:2000bw}.

A possible strategy for extending these benchmark results to the so far inaccessible regions of the QCD phase diagram is to use effective theories for the low-energy sector of QCD, which reproduce lattice results at vanishing and small $\mu$, and systematically extend the predictions to high chemical potential without
changing the model inputs fixed with lattice results for the QCD vacuum.
That leaves us with a variety of possibilities for the phase structure at nonzero $\mu$, depending on the effective model.

Among them we want to mention the following ones
\begin{itemize}
\item no CEP at all \cite{Bratovic:2012qs}, since the transition is
crossover in the whole phase diagram,
\item no CEP, but a Lifshitz point \cite{Carignano:2010},
\item one CEP, but with largely differing predictions of its position
\cite{Stephanov:2007fk},
\item second CEP \cite{Kitazawa:2002bc,Blaschke:2003cv,Hatsuda:2006ps,Bowman:2008kc},
\item several CEPs \cite{Kunihiro:2010vh,Zhang:2009mk},
\item CEP and triple point, possibly coincident, due to another phase
(i.e. colour superconducting \cite{Blaschke:2004cc} or quarkyonic matter \cite{Andronic:2009gj}) at low temperatures and high densities.
\end{itemize}

This spectrum of possibilities is rather broad in view of the upcoming
experimental programmes.
It is crucial to analyze the predictions arising from effective
models for their compatibility with lattice results.

One of the effective models that accounts for dynamical breaking of chiral
symmetry and its restoration at finite $T$ and $\mu$ is the Nambu-Jona-Lasinio
(NJL) model \cite{Nambu,Vogl:1991qt,Klevansky:1992qe,Hatsuda:1994pi}.
The absence of confinement in this model is partially cured by coupling its
chiral quark sector to the Polyakov-loop variable and adjusting a suitable
potential with a temperature dependence that is adjusted to describe the
pressure in accordance with lattice QCD simulations in the pure gauge field
system \cite{Ratti:2005jh,Roessner:2006xn,Sasaki:2006ww}.
These PNJL models provide a straightforward approach to the behaviour of
chiral and Polyakov-loop order parameters in the $T-\mu$ plane
(the phase diagram, see also \cite{Fukushima2008,Abuki:2008nm})
and predict a position for the CEP.

It has been shown that the non-local version of the PNJL model,
reproduces hadron properties at zero density and temperature and presents some
advantages over the local model
\cite{Schmidt:1994di,Efimov:1995uz,Contrera:2007wu}.
As another feature, one can add to the model a vector repulsive interaction
which increases the stiffness of quark matter and is therefore indispensable
to discuss under the observational constraint of $2~M_\odot$ neutron stars
\cite{Demorest:2010bx,Antoniadis:2013pzd}
the possibility of quark matter phases in their interiors
\cite{Klahn:2006iw,Orsaria:2012je,Klahn:2013kga,Orsaria:2013hna,Shao:2013toa}.
Such astrophysical applications have recently also been considered within the
nonlocal PNJL model \cite{Blaschke:2007ri,Blaschke:2013rma,Blaschke:2013ana,Alvarez-Castillo:2013spa}.

Our aim in the present paper is to make a systematic study of the
location of the CEP based on chiral quark models constrained from lattice
results, including all interactions mentioned above.
We tune the parameters of our model to reproduce lattice results at zero
density and then we extrapolate our predictions to regions of finite density
or chemical potential.

This article is organized as follows.
In Sect. II we present the description of the model and the parametrizations
we used. In Sect. III we show our results for the different form factors and
parameters. Then, in Sect. IV we present our conclusions.

\section{General Formalism}
Let us start describing the general formalism of the model we used. In the present work we considered a non local $SU(2)_f$ chiral model, including vector interactions as well as quark couplings to the gauge color background fields.
\subsection{Nonlocal chiral quark model}
The corresponding Lagrangian of the model used in this work is given by
\begin{equation}
{\cal {L}} = \bar{q}(i\slashed{D}-m_0) q + {\cal L}_{\rm int} +
{{\cal U}}(\Phi)~,
\label{lagr}
\end{equation}
where $q$ is the $N_{f}=2$ fermion doublet $q\equiv(u,d)^T$,
and $m_0$ is the current quark mass (we consider isospin symmetry, that is
$m_0=m_{u}=m_{d}$). The covariant derivative is defined as
$D_\mu\equiv \partial_\mu - iA_\mu$, where $A_\mu$ are color gauge fields.

The nonlocal interaction channels are given by
 \begin{equation}
{\cal L}_{\rm int}= -\frac{G_{S}}{2} \Big[ j_{a}(x)j_{a}(x)- j_{P}%
(x)j_{P}(x)\Big] {-}
\frac{G_V}{2} j_V(x)\, j_V(x),
\label{lagrangian}
\end{equation}
where the nonlocal currents are
\begin{eqnarray}
j_{a}(x)  &  =&\int d^{4}z\ g(z)\ \bar{q}\left(x+\frac{z}{2}\right)
\ \Gamma_{a}\ q\left(  x-\frac{z}{2}\right)  \ ,\nonumber\\
j_{P}(x)  &  =&\int d^{4}z\ f(z)\ \bar{q}\left(x+\frac{z}{2}\right)
\frac{i {\overleftrightarrow{\rlap/\partial}}}{2\ \kappa_{p}}
\ q\left( x-\frac{z}{2}\right) \ ,\nonumber\\
j_{V}(x)  &  =&\int d^4 z \ g(z)\ \bar{q}\left(x+\frac{z}{2}\right)
\,\gamma^0 \ q\left( x-\frac{z}{2}\right)
\label{eq:currents}
\end{eqnarray}
with $\Gamma_{a}=(\Gamma_{S},\Gamma_{P})=(\leavevmode\hbox{\small1\kern-3.8pt\normalsize1},i\gamma
_{5}\vec{\tau})$ for scalar and pseudoscalar currents respectively, and
$u(x^{\prime}){\overleftrightarrow{\partial}%
}v(x)=u(x^{\prime})\partial_{x}v(x)-\partial_{x^{\prime}}u(x^{\prime})v(x)$.
The functions $g(z)$ and $f(z)$ are
nonlocal covariant form factors. The scalar-isoscalar component of the $j_{a}(x)$
current is the responsible for the momentum dependence of the quark mass in the
quark propagator. Then, the current $j_{P}(x)$ will generate a momentum dependent wave function renormalization (WFR) of this propagator.
The mass parameter $\kappa_{p}$ controles the relative strength between both interaction terms in (\ref{eq:currents}).
Finally, $j_{V}(x)$ represents the vector channel interaction current, whose
coupling constant $G_V$ is usually taken as a free parameter. Moreover, we also have considered in this vector interaction the same nonlocal covariant form factor $g(z)$ used for the scalar and pseudoscalar currents. Then it is not necessary to include new free parameter in this term.

After the Fourier transform into momentum space, we have performed a standard bosonization of the theory introducing the bosonic fields $\sigma_{1,2}(p)$ and $\omega(p)$ and integrate out the quark fields. Furthermore, as we work within the mean field approximation (MFA), we replace the bosonic fields by their vacuum expectation values $\sigma_{1,2}$ and $\omega$ respectively, and the corresponding fluctuations are neglected.
The main motivation of the present work is to study the phase diagram of the strongly interacting quark matter. As we want to analyze the chiral phase transition for different choices of the parameters of the model, we have to include the dependence on the temperature $T$ and quark chemical potential $\mu$ in our effective action. \footnote {The corresponding values for baryon chemical potential $\mu_B$ can be easily obtained from the relation $\mu_B = 3~\mu$.}
In the present work this is carried out by using the Matsubara imaginary time
formalism. As mentioned above, the quarks are coupled to the gluons in (\ref{lagr}) through the covariant derivative.

The coupling of fermions with the gluon fields is taken into account in (\ref{lagr})
through the covariant derivative. Considering the quarks in a color field background
$A_0 = g δ_{µ0} G^a_{\mu} \lambda_a /2$, where $G^a_{\mu}$ are the SU(3) color gauge fields. At mean field level the traced Polyakov loop is given by $\Phi=\frac{1}{3} {\rm Tr}\,\exp( i\beta \phi)$, with $\phi = i\bar A_0$. Then, in the Polyakov gauge, the matrix $\phi$ is given by $\phi = \phi_3 \lambda_3 + \phi_8 \lambda_8$.
Considering that the mean field expectation values of $\Phi$ must be real, we set $\phi_8 = 0$ \cite{Roessner:2006xn,Contrera:2007wu}. Then, the mean field traced Polyakov loop is then given by $\Phi = \Phi^* = [1 + 2 $cos$(\phi_3/T)]/3$. Finally, the lagrangian (\ref{lagr}) also includes an effective potential ${\cal U}$ that accounts for gauge field self-interactions and will be briefly described below.

Within this framework the mean field thermodynamical potential
$\Omega^{\rm MFA}$ results
\begin{eqnarray}
\Omega^{\rm MFA} =  \,&-& \,{4 T} \sum_{c} \sum_n \int \frac{d^3\vec p}{(2\pi)^3} \, \,
\mbox{ln} \left[ \frac{ ({\rho}_{n, \vec{p}}^c)^2 +
M^2(\rho_{n,\vec{p}}^c)}{Z^2(\rho_{n, \vec{p}}^c)}\right] \nonumber \\
&+&
\frac{\sigma_1^2 + \kappa_p^2\ \sigma_2^2}{2\,G_S} - \frac{\omega^2}{2 G_V}\ + {{\cal U}}(\Phi ,T) \label{granp}
\end{eqnarray}
where $M(p)$ and $Z(p)$ are given by
\begin{eqnarray}
M(p) & = & Z(p) \left[m + \sigma_1 \ g(p) \right] ,\nonumber\\
Z(p) & = & \left[ 1 - \sigma_2 \ f(p) \right]^{-1}.
\label{mz}
\end{eqnarray}
Finally, as in \cite{Contrera:2010kz} we have considered
\begin{equation}
\Big({\rho_{n,\vec{p}}^c} \Big)^2 =
\Big[ (2 n +1 )\pi  T - i \mu + \phi_c \Big]^2 + {\vec{p}}\ \! ^2 \ ,
\label{eq:rho}
\end{equation}
where $\phi_c$ are given by the relation $\phi =
{\rm diag}(\phi_r,\phi_g,\phi_b)$. Namely, $\phi_c = c \ \phi_3$
with $c = 1,-1,0$ for $r,g,b$, respectively.

For finite values of the current quark mass, $\Omega^{\rm MFA}$ turns out to be divergent. The regularization procedure used here considers \cite{Contrera:2010kz,Dumm:2005}

\begin{equation}
\Omega^{\rm MFA}_{(reg)} = \Omega^{\rm MFA} - \Omega^{free}
+ \Omega^{free}_{(reg)},
\label{omegareg}
\end{equation}
where $\Omega^{free}$ is obtained from (\ref{granp}) for $\sigma_1 = \sigma_2 = 0 $, and $\Omega^{free}_{(reg)}$ is the regularized expression for
the thermodynamical potential of a free fermion gas,
\begin{equation}
\Omega^{free}_{(reg)} = -4\ T \int \frac{d^3 \vec{p}}{(2\pi)^3}\;
\sum_{c} \left[ \ln\left( 1 + e^{-\left( \sqrt{\vec{p}^2+m^2 }-\mu +i \phi_c
\right)/T} \right) + \ln\left( 1 + e^{-\left(
\sqrt{\vec{p}^2+m^2}+\mu + i \phi_c \right)/T} \right) \right].
\label{freeomegareg}
\end{equation}

Next step is obtaining the mean field values of $\sigma_1 ,\, \sigma_2 ,\, \omega $ and $\Phi$ as a function of the chemical potential and the temperature, by solving the following set of coupled equations

\begin{equation}
\frac{ d \Omega^{\rm MFA}_{(reg)}}{d \sigma_1} = 0
\qquad , \qquad
\frac{ d \Omega^{\rm MFA}_{(reg)}}{d \sigma_2} = 0
\qquad , \qquad
\frac{ d \Omega^{\rm MFA}_{(reg)}}{d \omega} = 0
\qquad , \qquad
\frac{ d \Omega^{\rm MFA}_{(reg)}}{d \Phi} = 0
\ .
\end{equation}

Then we can evaluate the $\Omega^{\rm MFA}_{(reg)}(\sigma_1,\sigma_2,\omega,\Phi,\mu,T)$ and compute all the relevant thermodynamic quantities needed in our calculation, like the chiral quark condensate

\begin{equation}
\langle \bar \psi \psi \rangle = \frac{ \partial \Omega^{\rm MFA}_{(reg)}}{\partial m}
\qquad , \qquad
\rho_q = - \frac{ \partial \Omega^{\rm MFA}_{(reg)}}{\partial \mu}\ .
\end{equation}
and the chiral susceptibility $\chi$ which can be used to determine the
characteristic of the chiral phase transition

\begin{equation}
\chi_{ch} = - \frac{ \partial^2 \Omega^{\rm MFA}_{(reg)}} {\partial m^2} =
- \frac{ \partial \langle \bar \psi \psi \rangle} {\partial m}\ .
\end{equation}

To proceed, we still have to define some quantities like the form factors, the vector interactions, the effective potential ${\cal U}$ and as well the parameters of the model. Let us introduce them gradually.

\subsection{Form factors and wave function renormalization}

Following \cite{Contrera:2010kz,Noguera:2008prd78}, we have considered two different types of functional dependency for the form factors $g(q)$ and $f(q)$:

exponential forms
\begin{eqnarray}
(\textrm{Set A}) \quad \left\{
\begin{array}{l}
g(p)= \mbox{exp}\left(-p^{2}/\Lambda_{0}^{2}\right) \\
f(p)=0
\end{array}
\right.
\label{eq:setA}
\end{eqnarray}
\begin{eqnarray}
(\textrm{Set B}) \quad \left\{
\begin{array}{l}
g(p)= \mbox{exp}\left(-p^{2}/\Lambda_{0}^{2}\right) \\
f(p)= \mbox{exp}\left(-p^{2}/\Lambda_{1}^{2}\right)
\end{array}
\right.
\label{eq:setB}
\end{eqnarray}
and Lorentzians with WFR
\begin{eqnarray}
(\textrm{Set C}) \quad \left\{
\begin{array}{l}
g(p)  = \frac{1+\alpha_z}{1+\alpha_z\ f_z(p)} \frac{\alpha_m \ f_m(p) -m\ \alpha_z f_z(p)}
{\alpha_m - m \ \alpha_z } \\
f(p)  = \frac{ 1+ \alpha_z}{1+\alpha_z \ f_z(p)} f_z(p)\
\end{array}
\right.
\label{eq:setC}
\end{eqnarray}
where
\begin{equation}
f_{m}(p) = \left[ 1+ \left( p^{2}/\Lambda_{0}^{2}\right)^{3/2} \right]^{-1}
\qquad , \qquad
f_{z}(p) = \left[ 1+ \left( p^{2}/\Lambda_{1}^{2}\right) \right]^{-5/2},
\label{eq:setC2}
\end{equation}
and $\alpha_m = 309$ MeV, $\alpha_{z}=-0.3$. All the parameter sets are summarized in the table (\ref{tab:parameters}).

\begin{table}[htb]
\begin{center}
 \begin{tabular}{|c|c|c|c|}
   \hline
   -- & Set A & Set B & Set C \\ \hline
   m [MeV] & 5.78 & 5.7 & 2.37 \\
   $\Lambda_0$ [MeV] & 752.2 & 814.42 & 850.0\\
   $G_S\, \Lambda_0^2$&  $20.65$ & $32.03$ & $20.818 $ \\
   $\Lambda_1$ [MeV] & 0.0 & 1034.5 & 1400.0 \\
   $\kappa_p$ [MeV] & 0.0 & 4180.0 & 6034.0 \\
   \hline
 \end{tabular}
\end{center}
\caption{ Sets of parameters (see Ref. \cite{Noguera:2008prd78} for a detailed description).}
\label{tab:parameters}
\end{table}

\begin{figure}[hbtp]
\begin{center}
\includegraphics[scale=0.2]{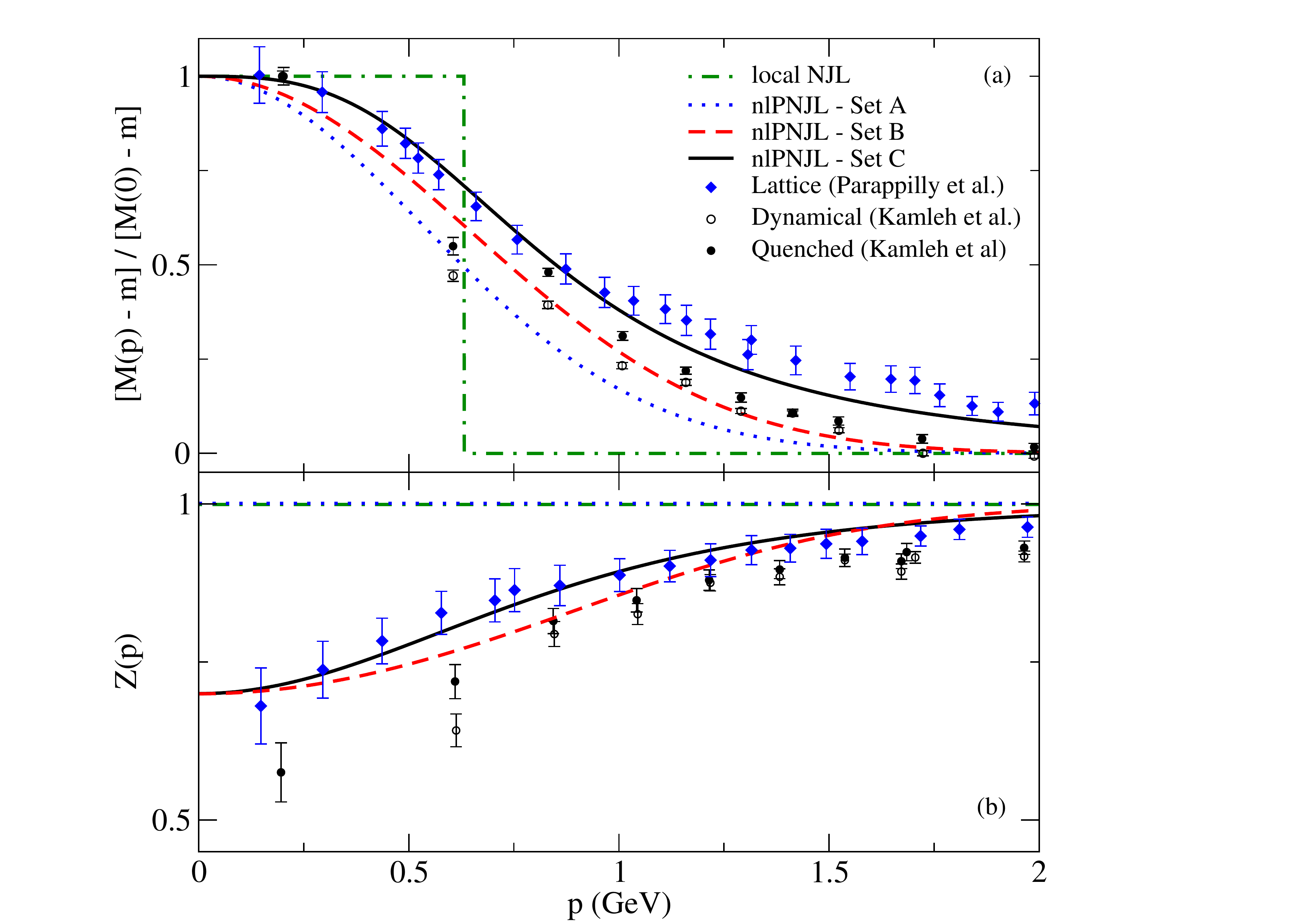}
\end{center}
\caption{(Color online) Normalized dynamical masses (a) and wave function renormalization (b) for the different non local form factors under study from \cite{Contrera:2010kz,Noguera:2008prd78} fitted to lattice data from \cite{Parappilly:2005ei}. For comparison the local model \cite{GomezDumm:2008sk} and more recent lattice data from \cite{Kamleh:2007ud} are also included.}
\label{fig:1}
\end{figure}

In addition, we want to include in our analysis the results arising from a local PNJL
model, what allow us to compare with the results obtained for example in \cite{Bratovic:2012qs}.
For that purpose, we started from the Lagrangian in \cite{Fukushima2008} with
two flavors instead of three and we use the same model parameters as in Refs. \cite{Ratti:2005jh,Roessner:2006xn,GomezDumm:2008sk}:
\begin{equation}
m=5.5~\textrm{MeV}; \; G=10.1~\textrm{GeV}^{-2}; \; \Lambda= 650.0~\textrm{MeV}.
\end{equation}
The model inputs can be constrained with results from lattice QCD studies. In particular, the above described form factors of the nonlocal interaction have been chosen \cite{Noguera:2008prd78} such as to reproduce the dynamical mass function $M(p)$ and the WFR $Z(p)$ of the quark propagator in the vacuum \cite{Parappilly:2005ei}.
In Figure \ref{fig:1} we show the shapes of normalized dynamical masses and WFR for the models under discussion here, i.e., the nonlocal models with set A (rank-one), set B and set C (rank-two) parameterizations as well as the local limit.

This figure generalizes the corresponding Figure \ref{fig:1} in \cite{Noguera:2008prd78}, by including the local limit for $M(p)$ and showing the line $Z(p)=1$ for both, set A and the local case. For comparison, a more recent lattice data from \cite{Kamleh:2007ud} is also included. It is easy to recognize the better agreement between the lattice results and the more complete model, namely nonlocal PNJL with WFR, in its two form factor parameterizations given by set B and set C.

{\subsection{The Vector Coupling}

The vector coupling constant $G_V$ is considered a free parameter which in the mean field approximation (MFA) may be adjusted such as to reproduce the behavior
of the critical temperature, $T_c(\mu)$, which has recently been obtained by Taylor expansion technique in lattice QCD \cite{Kaczmarek:2011zz}
\begin{equation}
T_c(\mu)/T_c(0)= 1 - \kappa (\mu/T)^2 + {\mathcal{O}}[(\mu/T)^4],
\label{eq:lattice_cross}
\end{equation}
with $\kappa=0.059 (2) (4)$ being the curvature. Note that this result is not yet based on continuum extrapolated lattice results, so that discretization errors has to be expected. However, we are interested to present here a scheme for constraining effective QCD models. The quality of predictions can subsequently be improved by using better lattice QCD data and constant discretization schemes.In what follows, the vector coupling strength will be evaluated by considering different ratios of $\eta_V= G_V/G_S$ and we use this parameter to tune our model to obtain better agreement with lattice predictions.

We include the vector coupling as a shifting in the chemical potential according to
\begin{equation}
\tilde{\mu} = \mu \; - \omega \; g(p) \; Z(p).
\label{mutilde}
\end{equation}
Note that we include the WFR $Z(p)$ in the shifted chemical potential in order to keep the thermodynamical potential at mean field level.We found that the results are quite different if the $Z(p)$ had not been included into the shift. As an example of that, we show in Figure \ref{fig:2} the phase diagram for set B with and without WFR in $\tilde{\mu}$, for a particular value of $\eta_V$ (similar results were obtained within set C) and a finite value of vector coupling constant.

\begin{figure}[hbtp]
\centering
\includegraphics[scale=0.4]{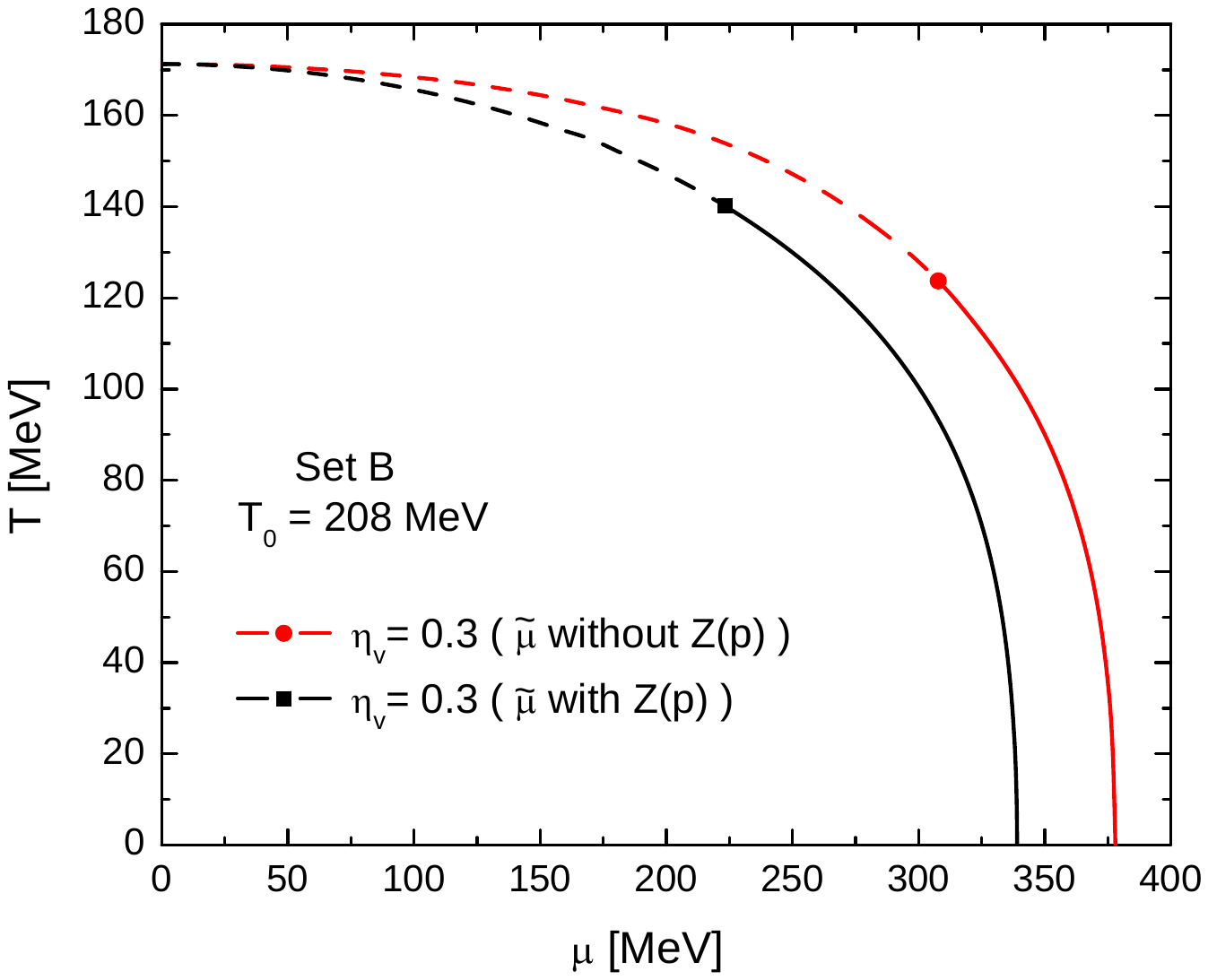}
\caption{(Color online) Comparative plot of phase diagram obtained with and without the inclusion of $Z(p)$ in the shifted chemical potential in equation \ref{mutilde}. Dashed lines represent Crossover transitions, the symbols (dot or square) indicate the critical end points location, and solid lines are first order phase transitions.}
\label{fig:2}
\end{figure}
Finally we have to include the shifting (\ref{mutilde}) in (\ref{eq:rho}) defining a new $\tilde{\rho}_{n,\vec{p}}^c$.

In the case of the local model, we consider that the chemical potential is shifted by
\begin{equation}
\tilde{\mu} = \mu \; - \omega.
\label{mutilde0}
\end{equation}


\subsection{Polyakov-loop potential}

In the present work we have chosen a $\mu$-dependent logarithmic effective potential described in \cite{Dexheimer:2009va}
\begin{equation}
{{\cal U}}(\Phi,T,\mu)=(a_0T^4+a_1\mu^4+a_2T^2\mu^2)\Phi^2
+ a_3T_0^4\ln{(1-6\Phi^2+8\Phi^3-3\Phi^4)},
\label{PL_pot}
\end{equation}
where $a_0=-1.85$, $a_1=-1.44$x$10^{-3}$, $a_2=-0.08$,
$a_3=-0.40$.

The reason for our particular election is that we want to consider the influence of chemical potential on the PL effective potential, and evaluate how this $\mu$-dependence can modify the phase diagram, tuning the vector coupling.
In the Figure \ref{fig:3} we compare the phase diagrams we obtained considering the $\mu$-dependent logarithmic effective potential and a non-$\mu$ dependent effective potential described in Ref.\cite{Roessner:2006xn}. The results for Set A and C have a similar qualitative behavior. As expected, at $T=0$ and $\mu=0$ both potentials produce the same critical temperatures, but there is a significant difference in the location of the CEPs. Nevertheless, it is shown in the same figure, that this difference turns to be smaller when increasing the vector coupling strength.

\begin{figure}[hbtp]
\begin{center}
\includegraphics[scale=0.4]{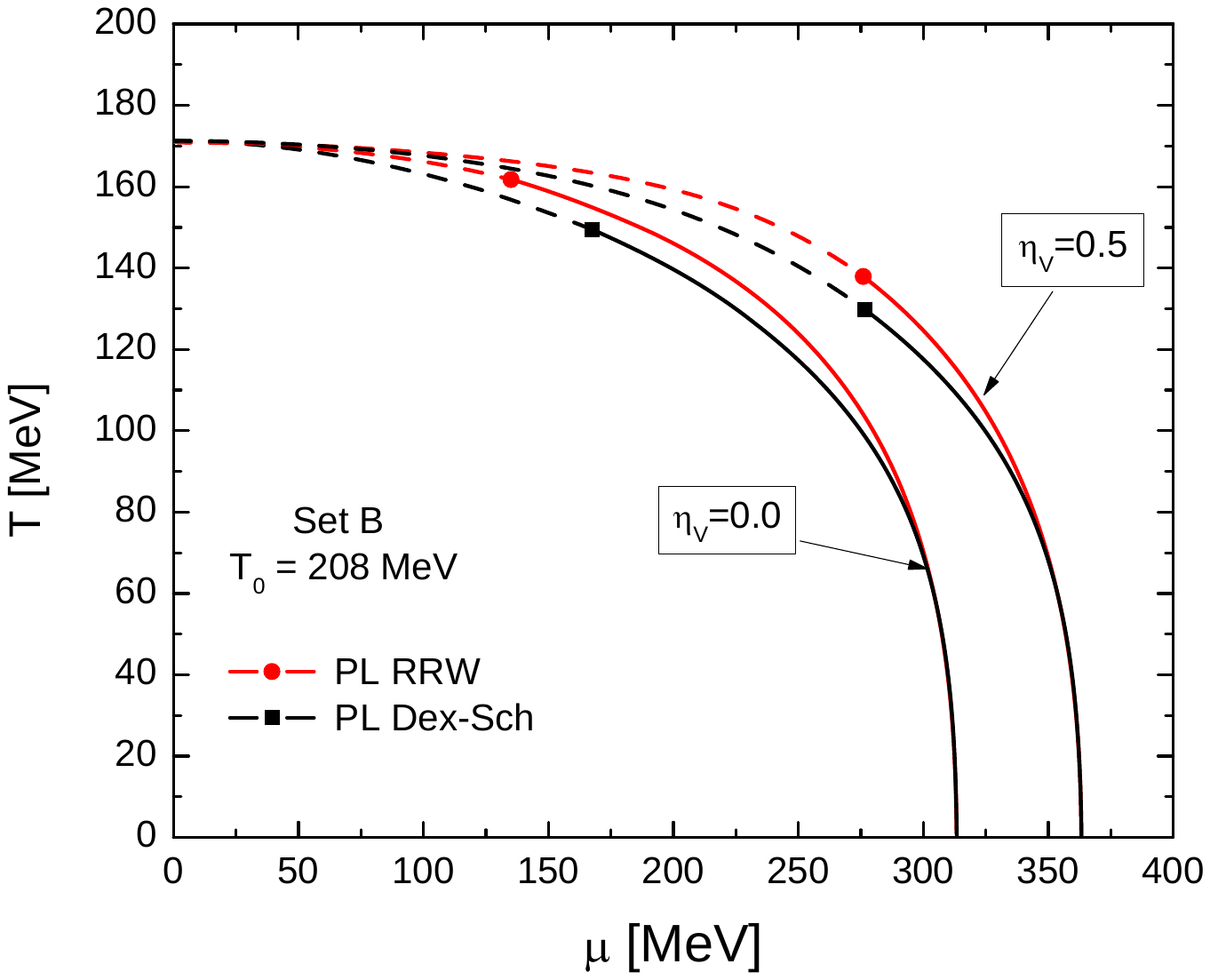}
\end{center}
\caption{(Color online) Phase diagrams obtained with the $\mu$-dependent effective potential from \cite{Dexheimer:2009va} used in this work in comparison with the logarithmic PL potential in Ref.\cite{Roessner:2006xn}. Dashed lines represent Crossover transitions, the symbols (dots or squares) indicate the critical end points location, and solid lines are first order phase transitions. }
\label{fig:3}
\end{figure}
Another point we have to discuss here is the election of the critical temperature $T_0$ for deconfinement transition. In the present work we set $T_0$ for deconfinement by using the value corresponding to two flavors, i.e. $T_0= 208$ MeV, as it has been suggested in ~\cite{Schaefer:2007pw}, and used in subsequent approaches, including the nonlocal PNJL ~\cite{Pagura:2012}, Polyakov loop-DSE models ~\cite{Horvatic:2010md} and entanglement PNJL (EPNJL) model ~\cite{Sakai:2010rp,Dutra:2013lya}. This election produces lower $T_c$ values for the chiral transition (see Figure \ref{fig:4} as an example), obtaining at zero chemical potential closer values to the lattice QCD more accepted result ~\cite{Ejiri:2000bw} for the critical temperature $T_c(0)=170$ MeV.
It is remarkable that within the local model the obtained critical temperatures are noticeably higher than in the case of nonlocal models, as it is shown in Figure \ref{fig:4}.

In this context it is important to keep in mind the effect that the account
for a hadronic phase can have on the topology of the QCD phase diagram
\cite{Blaschke:2010ka,Shao:2011fk}, as well as higher order quark interactions
\cite{Sasaki:2010jz}.

\begin{figure}[hbtp]
\begin{center}
\includegraphics[scale=0.35]{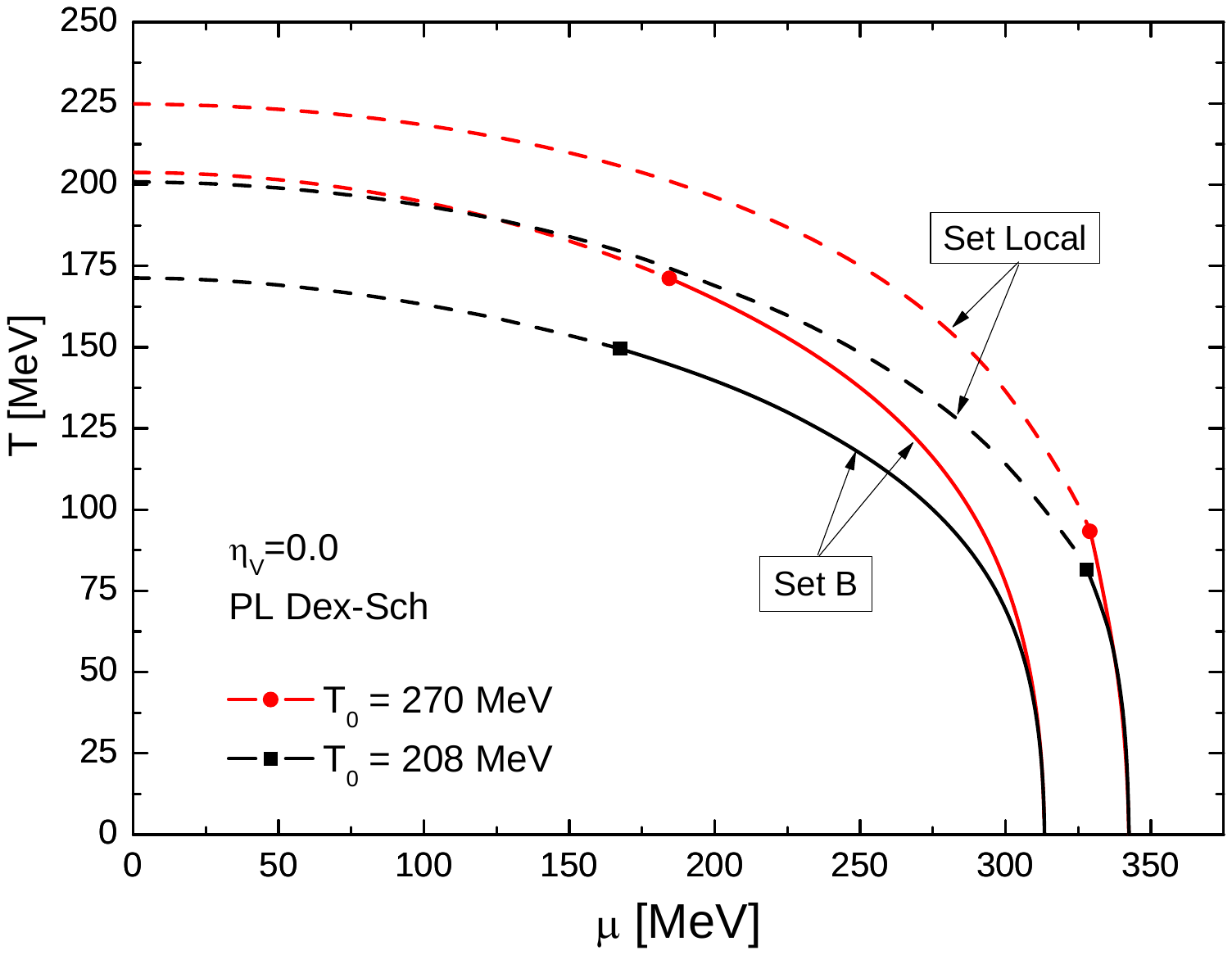}
\end{center}
\caption{(Color online) Comparison of phase diagrams determined with two different values of the deconfinement critical temperature $T_0$: the pure gauge value ($T_0=270$ MeV) and the two flavor value ($T_0=208$ MeV). See Refs. ~\cite{Schaefer:2007pw,Pagura:2012,Horvatic:2010md} for details. Dashed lines represent Crossover transitions, the symbols (dots or squares) indicate the critical end points location, and solid lines are first order phase transitions.}
\label{fig:4}
\end{figure}

\section{Results}
%
The first effect we want to study is how the vector interactions affect the transitions and the location of the CEP. We built the corresponding phase diagrams for different values of $\eta_V$. In all the cases we observed a variation in the curvatures even at low chemical potential. As expected, the influence of the vector coupling increases with the chemical potential, then the position of the CEP and the values of $\mu_c(T=0)$ reflect notably this influence. We want to remark that, for increasing $\eta_V$ the CEPs tend to be located towards lower $T$ and higher $\mu$. Similarly to what has been shown in Ref.\cite{Hell:2012da}, we observe that for any the nonlocal models under study, the CEPs (and the corresponding first order transitions) are still present for all the values of vector coupling constant analyzed in this work. In Figure \ref{fig:5} it can be seen that effect for Set B (cualitatively similar behavior was observed for set A and set C.). Nevertheless, for the local model, we observe that increasing $\eta_V$ a crossover line without a CEP is obtained, as it is shown in Figure \ref{fig:6}.

\begin{figure}[hbtp]
\centering
\includegraphics[scale=0.4]{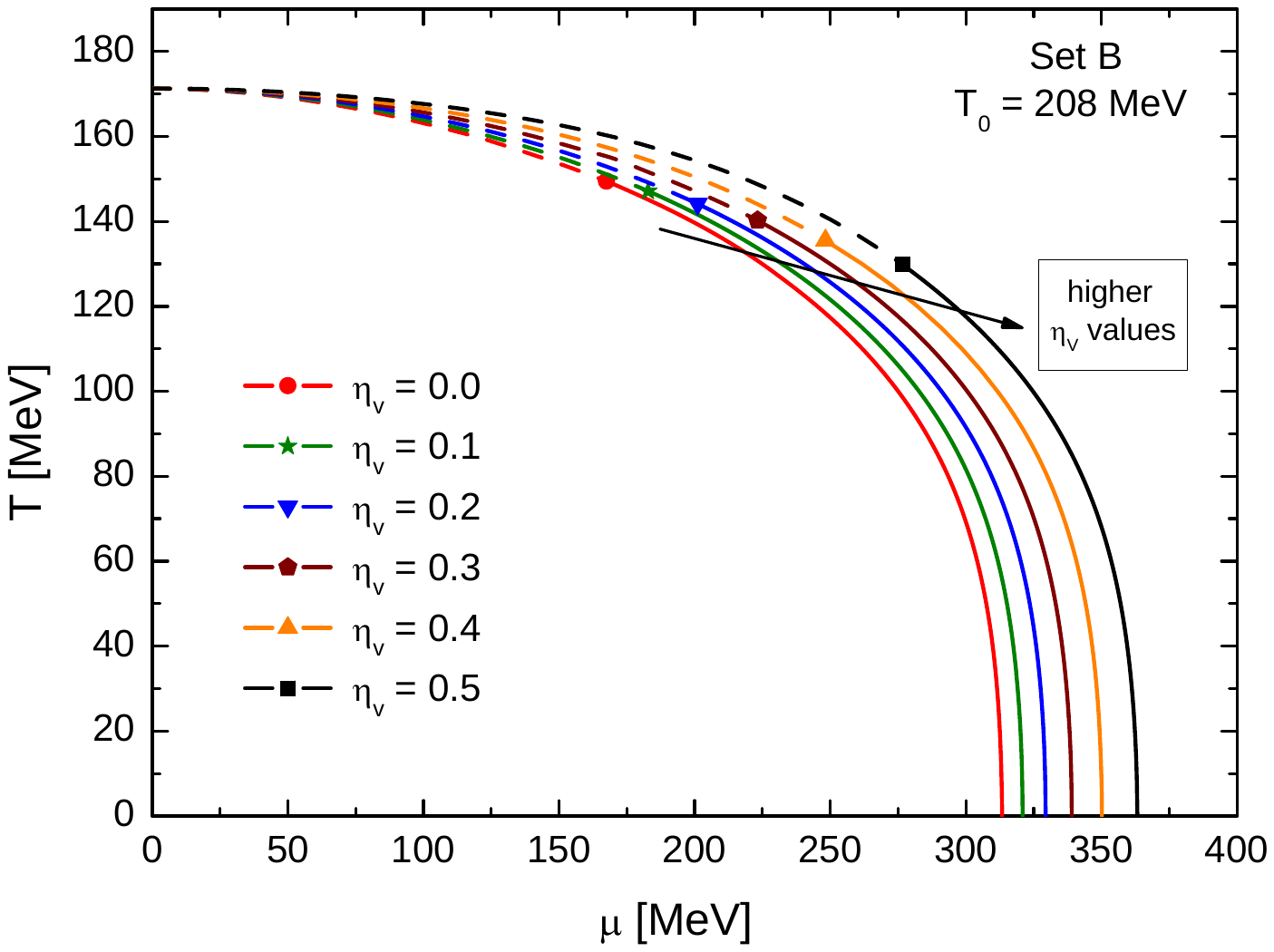}
\caption{(Color online) Influence of the strength of the vector coupling (here $\eta_V= G_V/G_S$) on the phase diagram for set B. Dashed lines represent Crossover transitions, the symbols indicate the critical end points locations, and solid lines are first order phase transitions.}
\label{fig:5}
\end{figure}

\begin{figure}[hbtp]
\centering
\includegraphics[scale=0.4]{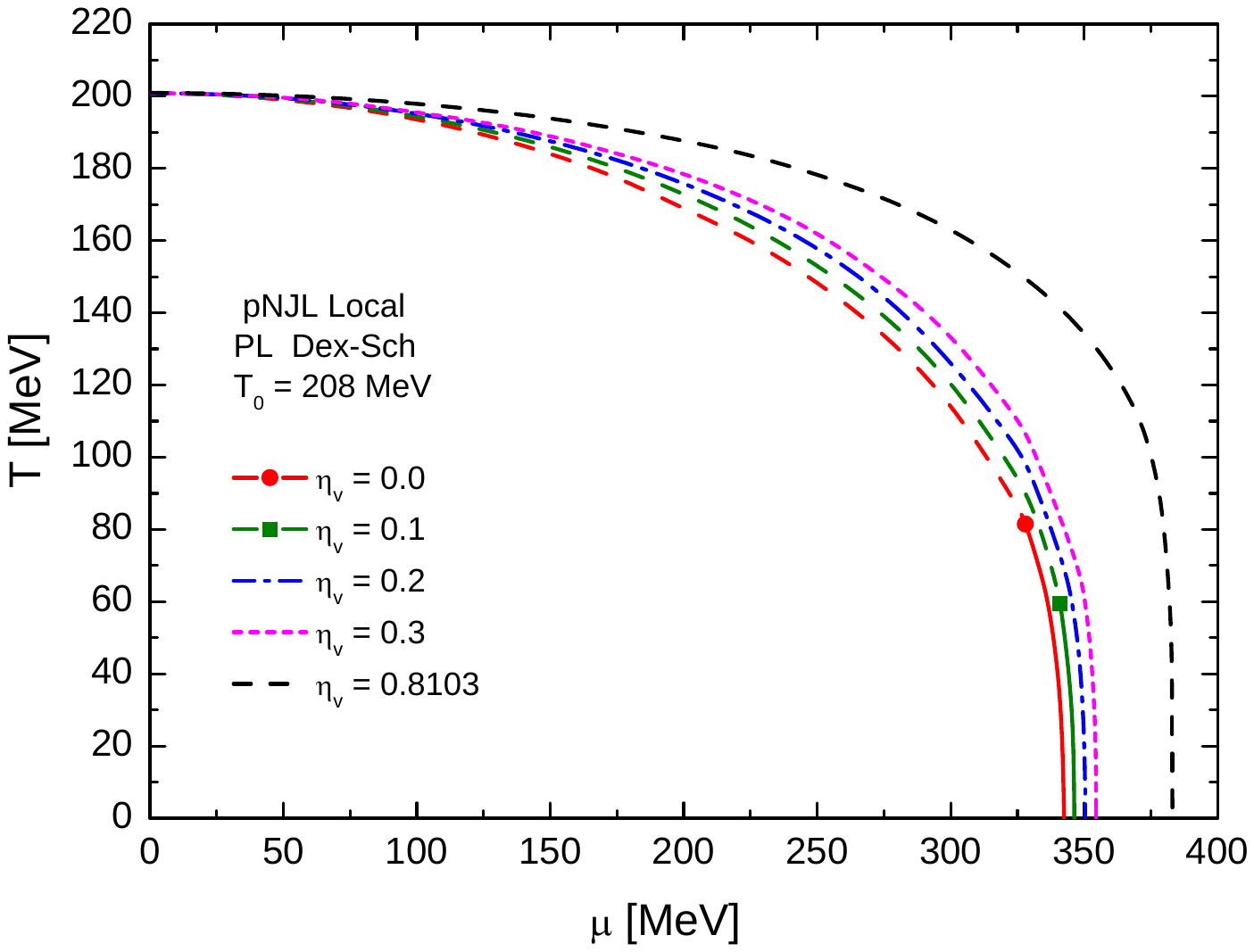}
\caption{(Color online) Influence of the strength of the vector coupling on the phase diagram for the local model. Dashed lines represent Crossover transitions, the symbols (dot or square) indicate the critical end points location, and solid lines are first order phase transitions. It can be seen that CEPs tends to disappear for $G_V>0.1~G_S$, leaving only a crossover phase transition.}
\label{fig:6}
\end{figure}

Once we have the phase diagrams, the next step is to determine the curvatures. To do so, we plotted the pseudocritical temperatures of the crossover transitions as function of $(\mu/T)^2$ for different $\eta_V$ ratios. Then, the curvatures can be obtained from the slope of the straight lines in the region of low $(\mu/T)$ values.
An example of this is shown in Figure \ref{fig:7} for set C (the corresponding plots for the other sets are qualitatively very similar). The fit of the lattice QCD results (\ref{eq:lattice_cross}) is also
shown. The grey zone corresponds to the error in the coefficient $\kappa$ obtained in \cite{Kaczmarek:2011zz}.

\begin{figure}[hbtp]
\centering
\includegraphics[scale=0.4]{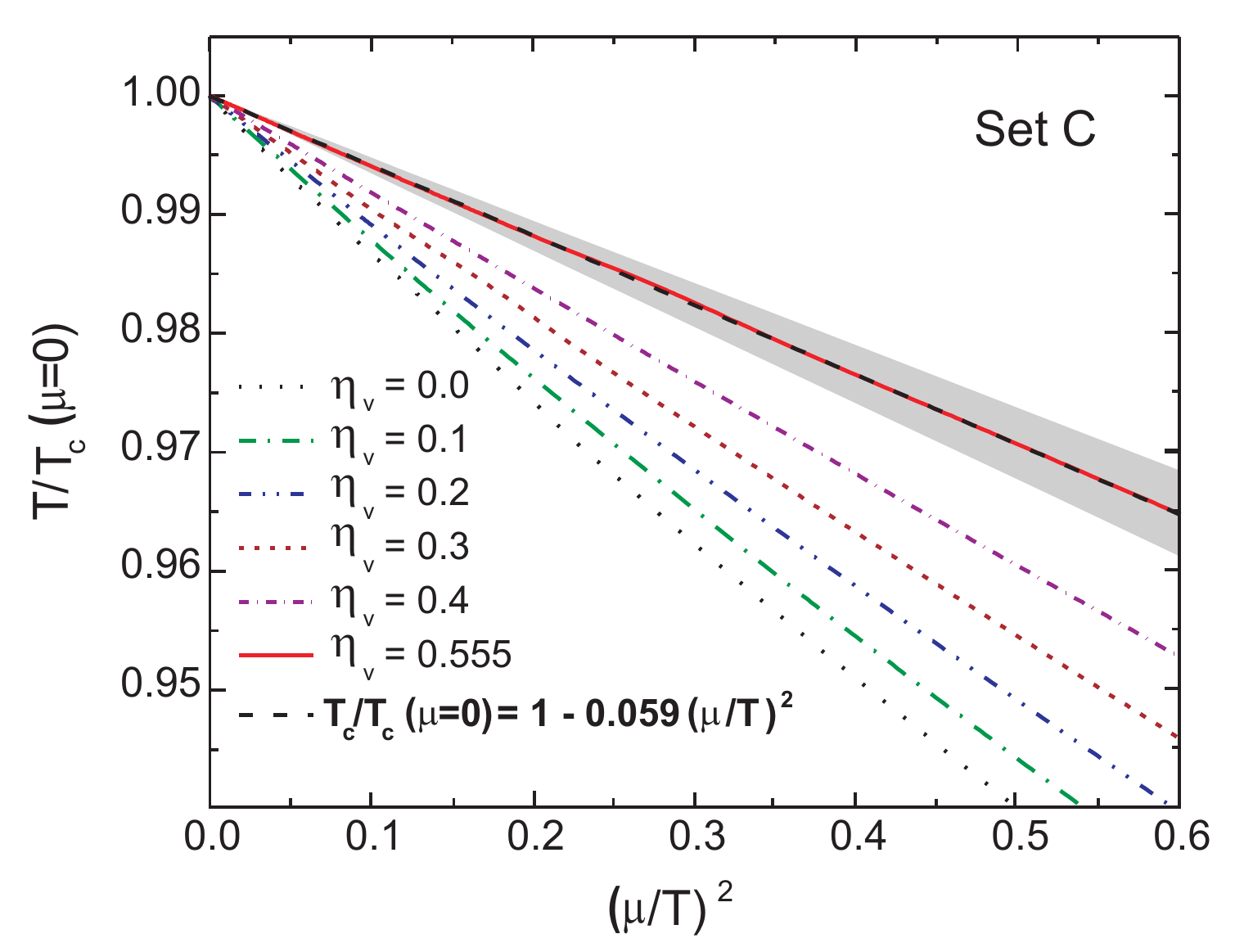}
\caption{(Color online) Chiral crossover transitions at low values of $\mu /T$ for different values of strengths of the vector coupling $\eta_V= G_V/G_S$ for Set C. The dashed line corresponds to the lattice QCD prediction of
$\kappa=0.059 (2) (4)$ \cite{Kaczmarek:2011zz}.}
\label{fig:7}
\end{figure}

In Figure \ref{fig:8} we compare the lattice result with the values for the
coefficient $\kappa$ obtained within the nonlocal PNJL models and the local one. There, the horizontal line corresponds to the lattice QCD prediction of $\kappa=0.059 (2) (4)$ \cite{Kaczmarek:2011zz} and the grey zone represents its error. Note that for the more complete model (set B and set C) the curvatures are closer to each other than in the case of set A and local ones.

It is important to remark that while the analysis of $\kappa$ has been performed for $N_f=2+1$ simulations, the chemical potential $\mu$ concerns only the two light flavors. Therefore, our extraction of $\kappa$ from the nonlocal PNJL models for the 2-flavor case may be in order.

\begin{figure}[hbtp]
\centering
\includegraphics[scale=0.4]{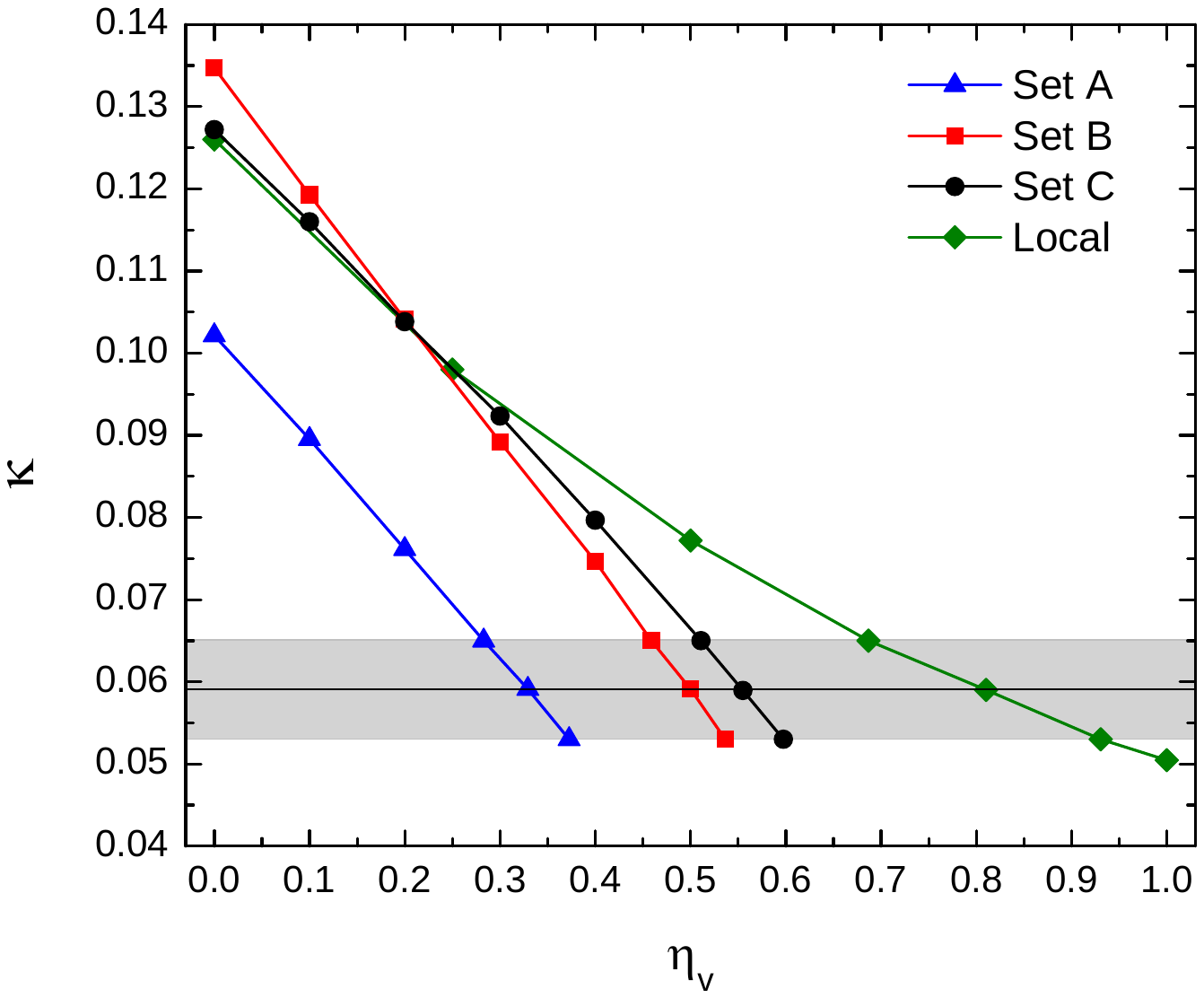}
\caption{(Color online) Curvature $\kappa$ of the pseudocritical temperature $T_c(\mu)$ of the
chiral crossover transition at low values of $\mu /T$.
The horizontal line corresponds to the lattice QCD prediction of
$\kappa=0.059 (2) (4)$ \cite{Kaczmarek:2011zz}.The grey zone represents the corresponding error in the curvature determination.}
\label{fig:8}
\end{figure}
In Figure \ref{fig:9} it is compared the phase diagrams for the nonlocal models discussed in this work, considering the corresponding $\eta_V$ values that best fit the lattice QCD prediction of
$\kappa=0.059 (2) (4)$ from \cite{Kaczmarek:2011zz}. The grey zones correspond to the range of $\eta_V$ values obtained by considering the error in the lattice determination of $\kappa$ \cite{Kaczmarek:2011zz}.
Similarly, the error bars in the CEP's indicate the distances to the CEP positions for the corresponding $\eta_V$ values that fit the error limits.

Note that using the Polyakov loop potential (\ref{PL_pot}) from \cite{Dexheimer:2009va}, a crossover region and a CEP can be obtained for set A, even for $T_0= 208$ MeV, contrarily to what has been reported in ~\cite{Pagura:2012,Horvatic:2010md}, where the Polyakov loop potential from \cite{Roessner:2006xn} has been used\footnote{Note that in the low-$\mu$ region the differences in the chiral transition obtained with each effective potential are mainly due to the definition of the corresponding logarithmic term in the Polyakov loop potential.}.

\begin{figure}[pbt]  
\includegraphics[width=158mm]{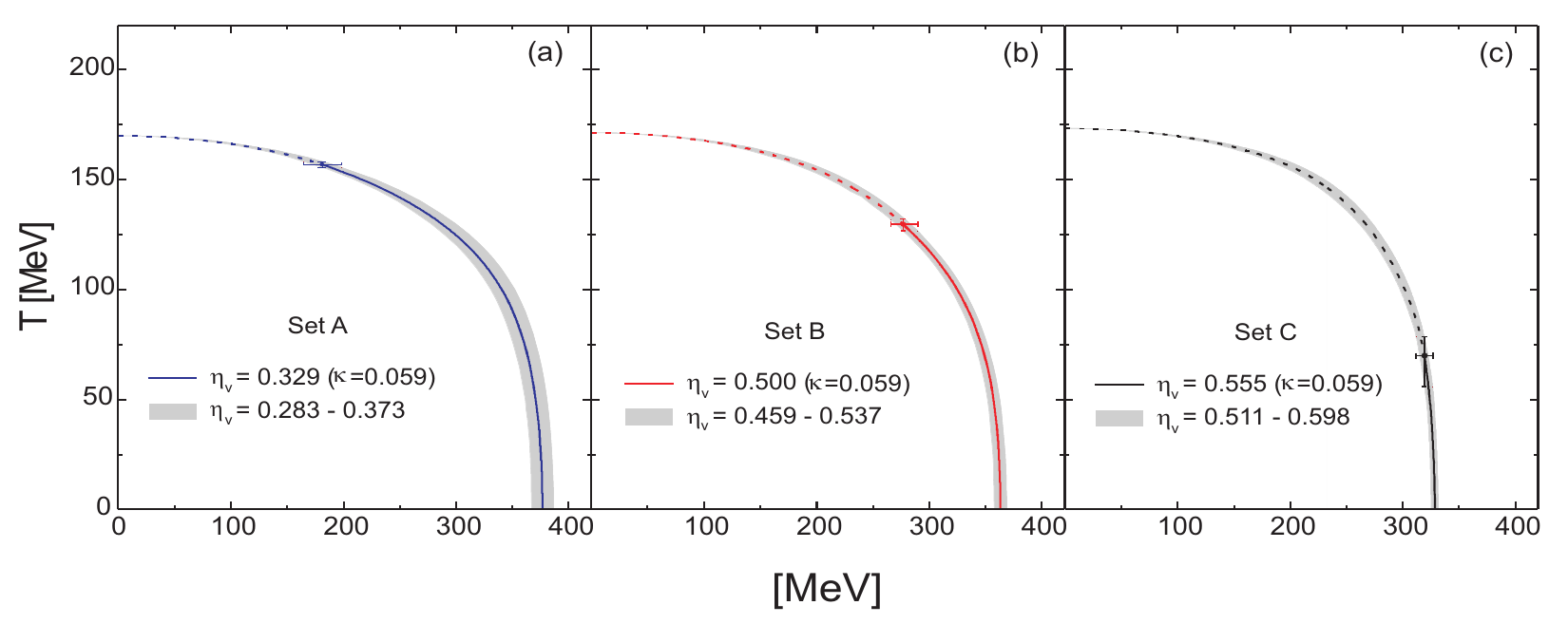}
\caption{ (Color online) Phase diagrams obtained with the non local models with the parametrizations Set A (a), Set B (b) and Set C (c), for the values of $\eta_V$ that best fit the lattice QCD prediction in
 \cite{Kaczmarek:2011zz}. The grey zones and the error bars of the CEP location represent the corresponding indetermination for the curvature value.}
\label{fig:9}
\end{figure}

The results summarized in Figure \ref{fig:10} indicate that the absolute value of the critical temperature $T_c(0)$ of nonlocal covariant PNJL models is rather insensitive to the choice of the form factors parametrizing, the momentum dependence of dynamical (running) mass function and WFR of the quark propagator as measured on the lattice at zero temperature \cite{Parappilly:2005ei}, whereas the position of the CEP and critical chemical
potential at $T=0$ strongly depends on it.
On the other hand, the value of $T_c(0)$ in the local model is significantly different (larger) than in the nonlocal ones. In addition, note that in the local model to fit the lattice QCD value requires a larger vector coupling, for which the corresponding phase diagram lacks of CEP and all the chiral phase transition is a crossover. This is another remarkable difference with respect to the obtained with non local models \cite{Bratovic:2012qs}.

\begin{figure}[hbtp]
\centering
\includegraphics[scale=0.4]{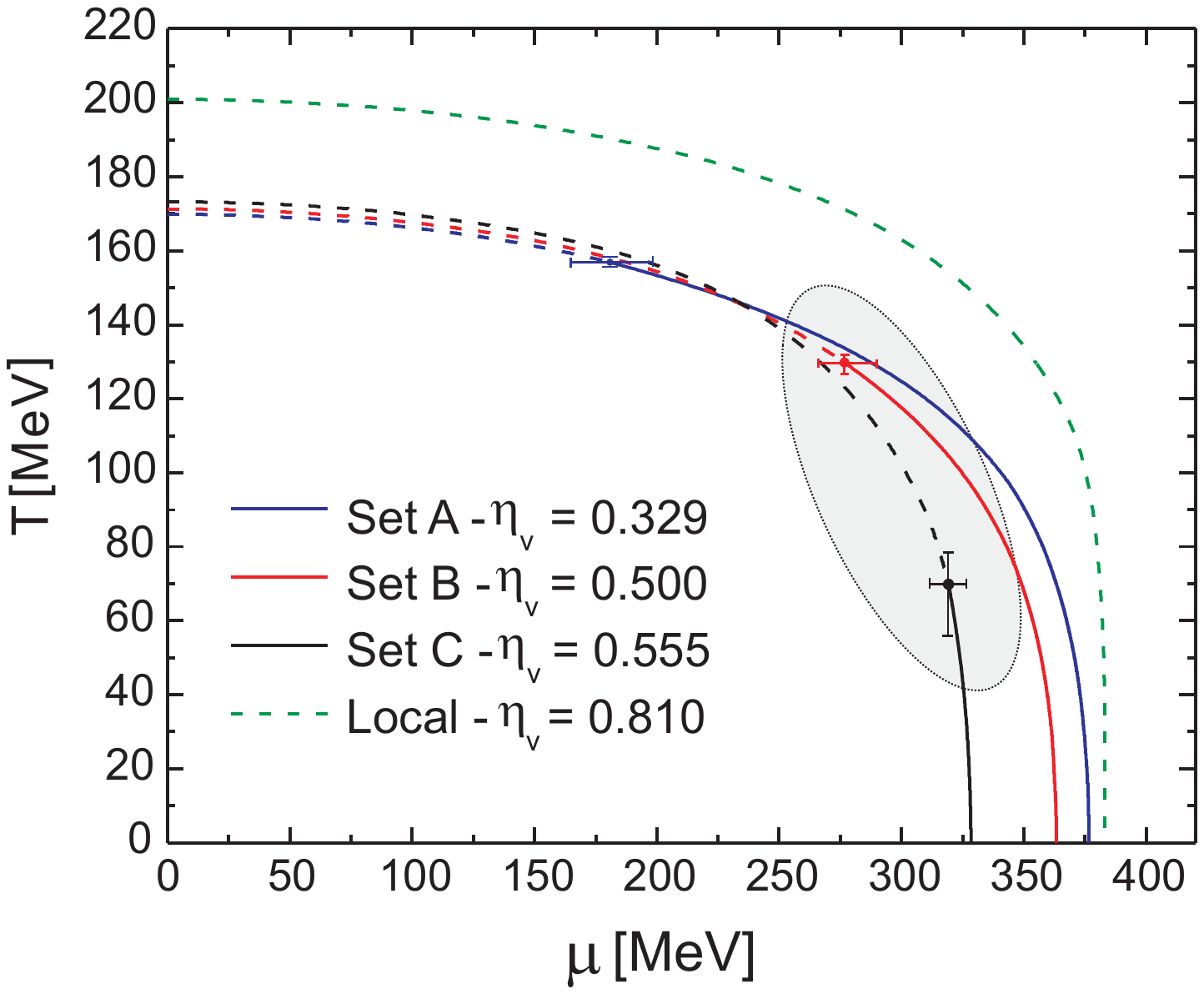}
\caption{(Colour online) Phase diagrams with (pseudo)critical temperatures $T_c(\mu)$ and
critical points for nonlocal PNJL models (set A-C) compared to the local one. Dashed (full) lines correspond to crossover (first order) transitions.
The vector coupling strength $\eta_V$ is chosen so that these models reproduce the lattice QCD result  $\kappa=0.059 (2)(4)$ \cite{Kaczmarek:2011zz} for the curvature at low $\mu$-values. The corresponding values for $T_c(\mu=0)$ (in MeV) are 169.9, 171.3, 173.2 and 200.9, for Sets A, B, C and local, respectively. The highlighted region denotes the CEP position favored by the present study.}
\label{fig:10}
\end{figure}

The nonlocal model for set B and set C cases, contains WFR and dynamical (running) quark mass effects and thus is closer to full QCD. Therefore, we suggest that statements about the existence and location of the CEP within set A and the local model should be less trustworthy than those of set B and set C.
As a consequence, a possible region for the CEP location suggested by our study would be between the results for set B and set C, i.e. around $(T_{\rm CEP},\mu_{\rm CEP})=(129.8~{\rm MeV},276.6~{\rm MeV})$ and
$(69.9~{\rm MeV},319.1~{\rm MeV})$, respectively.

This suggests that the search for CEP signatures in the BES programs is justified and should be continued. The energy range of the NICA and FAIR facilities shall be particularly promising.

\section{Summary and Conclusions}
In this work, based on nonlocal PNJL models with and without WFR, it is studied the influence of vector coupling strength in the QCD phase diagram. As shown if Figure \ref{fig:2}, it is observed a remarkable difference when the chemical potential shifting includes the WFR function, keeping the thermodynamical potential at mean field level.
In a further exploratory step, it is shown the influence on the phase diagrams of considering different values for the $T_0$ parameter in the effective potential (see Figure \ref{fig:4}). As expected it is observed a noticeable decrease of the chiral transition critical temperatures, but the smaller the temperature is the smaller is this decrease, converging to almost the same value when the critical temperature goes below $T=40$ MeV. Also in Figure \ref{fig:4} it can be seen a comparison between one of the nonlocal models (Set B as example) and the local one, where it is remarkable a difference of about 25 MeV along all the phase diagram lines for the same value of $T_0$. But the main variation is in the location of the CEP, which in the case of the local model turns to be at a lower $T$ and higher $\mu$ than the nonlocal case.
Comparing Figures \ref{fig:5} and \ref{fig:6} for nonlocal and local models respectively, it can be observed that the influence of the increasing of the vector coupling strength is qualitatively similar in both cases, i.e. moving the phase transitions towards higher $\mu$ values and lowering the CEP location. Nevertheless, in the case of the local model the CEP turns to disappear for higher values of $\eta_V$, leaving only a crossover transition along all the phase diagrams.

A second step in our work consisted in the determination of the curvatures in the low $\mu$ region as the slope of the straight lines obtained by plotting the pseudocritical temperatures of the crossover transitions as function of $(\mu/T)^2$. In Figure \ref{fig:7} it is shown some examples of the lines obtained with Set C and several values of $\eta_V$, while in Figure \ref{fig:8} it can be observed how the curvatures vary respect to the vector coupling strength for all the parametrizations considered.

Once it has been determined the values of $\eta_V$ that best reproduce the lattice QCD result $\kappa=0.059 (2)(4)$ \cite{Kaczmarek:2011zz} for the curvature at low $\mu$-values, we have shown (Figure \ref{fig:9}) the nonlocal model phase diagrams for the obtained values of $\eta_V$.

Finally, considering the more elaborated models (non local and with WFR, i.e. set B and set C), and the corresponding values of $\eta_V$ adjusted to lattice QCD results, we suggest in Figure \ref{fig:10} the most likely zone where, according to our study, the CEP would be located.

Note that in this exploratory study the lattice QCD data for the quark
propagator and for the curvature of the pseudocritical line are obtained
with different lattice actions. A more consistent study should be based on the
same discretization of the action, provide a continuum extrapolation and work
with physical quark masses.

As a next step it is necessary to investigate the robustness of the results
of the nonlocal PNJL models when modifying the choice of the Polyakov-loop
potential taking into account recent developments
\cite{Sasaki:2012bi,Ruggieri:2012ny,Fukushima:2012qa}
and, in particular, when going beyond the mean field approximation.
A scheme for going beyond the meanfield in nonlocal PNJL models by including
hadronic correlations (bound states and their dissociation in the continuum of
scattering states) has recently been developed \cite{Radzhabov:2010dd} and
shall be generalized for studies of the chiral and deconfinement phase
transition in the QCD phase diagram.
A key quantity for such studies will be the hadronic spectral function.
First results using a generic ansatz \cite{Blaschke:2011ry} for joining
the hadron resonance gas and PNJL approaches are promising
\cite{Turko:2011gw,Turko:2013taa} and have recently been underpinned a
microscopic justification by a treatment of pion dissociation within the
(local) PNJL model \cite{Wergieluk:2012gd}. This approach has recently been
generalized by mimicking confining properties with an infrared momentum cutoff
\cite{Dubinin:2013yga} in order to bind higher lying mesonic states such as
the sigma meson - a feature shared with the nonlocal extensions of the (P)NJL
model as, e.g., in \cite{Schmidt:1994di,Horvatic:2010md,Benic:2013tga},
to be exploited further in subsequent work.

\section*{Acknowledgements}
We would like to thank O. Kaczmarek, S. Nedelko, K. Redlich, C. Schmidt,
N. Scoccola for useful comments and discussions.
D.B. acknowledges hospitality and support during his visit at the University
of Bielefeld and funding of his research provided by the Polish Narodowe
Centrum Nauki within the ``Maestro'' programme under contract
UMO-2011/02/A/ST2/00306
as well as
by the Russian Fund for Basic Research under Grant No. 11-02-01538-a.
G.C. is grateful for support by CONICET (Argentina) and by ``CompStar'', a
research networking programme of the European Science Foundation.



\begin{thebibliography}{99}

\bibitem{Stephanov:2007fk}
  M.~A.~Stephanov,
  ``QCD phase diagram: An Overview,''
  PoS LAT {\bf 2006}, 024 (2006).

\bibitem{Bazavov:2011nk}
  A.~Bazavov, T.~Bhattacharya, M.~Cheng, C.~DeTar, H.~T.~Ding, S.~Gottlieb, R.~Gupta and P.~Hegde {\it et al.},
  ``The chiral and deconfinement aspects of the QCD transition,''
  Phys.\ Rev.\ D {\bf 85}, 054503 (2012).

\bibitem{Ejiri:2000bw}
  S.~Ejiri,
  ``Lattice QCD at finite temperature,''
  Nucl.\ Phys.\ Proc.\ Suppl.\  {\bf 94}, 19 (2001).

\bibitem{Bratovic:2012qs}
  N.~M.~Bratovic, T.~Hatsuda and W.~Weise,
  ``Role of Vector Interaction and Axial Anomaly in the PNJL Modeling of the QCD Phase Diagram,''
  Phys.\ Lett.\ B {\bf 719}, 131 (2013).

\bibitem{Carignano:2010}
  S.~Carignano, D.~Nickel, and M.~Buballa,
  ``Influence of vector interaction and Polyakov loop dynamics on inhomogeneous chiral symmetry breaking phases''
 Phys.\ Rev.\  D {\bf 82}, 054009 (2010).

\bibitem{Kitazawa:2002bc}
  M.~Kitazawa, T.~Koide, T.~Kunihiro and Y.~Nemoto,
  ``Chiral and color superconducting phase transitions with vector interaction in a simple model''
  Prog.\ Theor.\ Phys.\  {\bf 108}, 929 (2002).

\bibitem{Blaschke:2003cv}
  D.~Blaschke, M.~K.~Volkov and V.~L.~Yudichev,
  ``Coexistence of color superconductivity and chiral symmetry breaking within the NJL model,''
  Eur.\ Phys.\ J.\ A {\bf 17}, 103 (2003).

\bibitem{Hatsuda:2006ps}
  T.~Hatsuda, M.~Tachibana, N.~Yamamoto and G.~Baym,
  ``New critical point induced by the axial anomaly in dense QCD,''
  Phys.\ Rev.\ Lett.\  {\bf 97}, 122001 (2006).

\bibitem{Bowman:2008kc}
  E.~S.~Bowman and J.~I.~Kapusta,
  ``Critical Points in the Linear Sigma Model with Quarks,''
  Phys.\ Rev.\ C {\bf 79}, 015202 (2009).

\bibitem{Kunihiro:2010vh}
  T.~Kunihiro, Y.~Minami and Z.~Zhang,
  ``QCD Critical Points and Their Associated Soft Modes,''
  Prog.\ Theor.\ Phys.\ Suppl.\  {\bf 186}, 447 (2010).

\bibitem{Zhang:2009mk}
  Z.~Zhang and T.~Kunihiro,
  ``Vector interaction, charge neutrality and multiple chiral critical point structures,''
  Phys.\ Rev.\ D {\bf 80}, 014015 (2009).

\bibitem{Blaschke:2004cc}
  D.~Blaschke, H.~Grigorian, A.~Khalatyan and D.~N.~Voskresensky,
  ``Exploring the QCD phase diagram with compact stars,''
  Nucl.\ Phys.\ Proc.\ Suppl.\  {\bf 141}, 137 (2005).

\bibitem{Andronic:2009gj}
  A.~Andronic, D.~Blaschke, P.~Braun-Munzinger, J.~Cleymans, K.~Fukushima, L.~D.~McLerran, H.~Oeschler and R.~D.~Pisarski {\it et al.},
  ``Hadron Production in Ultra-relativistic Nuclear Collisions: Quarkyonic Matter and a Triple Point in the Phase Diagram of QCD,''
  Nucl.\ Phys.\ A {\bf 837}, 65 (2010).

\bibitem{Nambu}
Y.~Nambu and G.~Jona-Lasinio,
``Dynamical Model of Elementary Particles Based on an Analogy with
Superconductivity,''
Phys.\ Rev.\  {\bf 122}, 345 (1961); {\bf 124}, 246 (1961).


\bibitem{Vogl:1991qt}
  U.~Vogl and W.~Weise,
 ``The Nambu and Jona Lasinio model: Its implications for hadrons and nuclei,''
  Prog.\ Part.\ Nucl.\ Phys.\  {\bf 27}, 195 (1991).

\bibitem{Klevansky:1992qe}
  S.~P.~Klevansky,
  ``The Nambu-Jona-Lasinio model of quantum chromodynamics,''
  Rev.\ Mod.\ Phys.\  {\bf 64}, 649 (1992).

\bibitem{Hatsuda:1994pi}
  T.~Hatsuda and T.~Kunihiro,
  ``QCD phenomenology based on a chiral effective Lagrangian,''
  Phys.\ Rept.\  {\bf 247}, 221 (1994).

\bibitem{Ratti:2005jh}
  C.~Ratti, M.~A.~Thaler and W.~Weise,
  ``Phases of QCD: Lattice thermodynamics and a field theoretical model,''
  Phys.\ Rev.\ D {\bf 73}, 014019 (2006).

\bibitem{Roessner:2006xn}
  S.~Roessner, C.~Ratti and W.~Weise,
  ``Polyakov loop, diquarks and the two-flavour phase diagram,''
  Phys.\ Rev.\ D {\bf 75}, 034007 (2007).

\bibitem{Sasaki:2006ww}
  C.~Sasaki, B.~Friman and K.~Redlich,
  ``Susceptibilities and the Phase Structure of a Chiral Model with Polyakov Loops,''
  Phys.\ Rev.\ D {\bf 75}, 074013 (2007).

\bibitem{Fukushima2008}
  K.~Fukushima,
  ``Phase diagrams in the three-flavor Nambu-Jona-Lasinio model with the
  Polyakov loop,''
  Phys.\ Rev.\ D {\bf 77}, 114028 (2008);
  [Erratum-ibid.\ D {\bf 78}, 039902 (2008)].

\bibitem{Abuki:2008nm}
  H.~Abuki, R.~Anglani, R.~Gatto, G.~Nardulli and M.~Ruggieri,
  ``Chiral crossover, deconfinement and quarkyonic matter within a Nambu-Jona
Lasinio model with the Polyakov loop,''
  Phys.\ Rev.\ D {\bf 78}, 034034 (2008).


\bibitem{Schmidt:1994di}
  S.~M.~Schmidt, D.~Blaschke and Y.~.L.~Kalinovsky,
 ``Scalar - pseudoscalar meson masses in nonlocal effective QCD at finite temperature,''
  Phys.\ Rev.\ C {\bf 50}, 435 (1994).

\bibitem{Efimov:1995uz}
  G.~V.~Efimov and S.~N.~Nedelko,
  ``Nambu-Jona-Lasinio model with the homogeneous background gluon field,''
  Phys.\ Rev.\ D {\bf 51}, 176 (1995).

\bibitem{Contrera:2007wu}
  G.~A.~Contrera, D.~Gomez Dumm and N.~N.~Scoccola,
  ``Nonlocal SU(3) chiral quark models at finite temperature: the role of the
  Polyakov loop,''
  Phys.\ Lett.\  B {\bf 661}, 113 (2008).

\bibitem{Demorest:2010bx}
  P.~Demorest, T.~Pennucci, S.~Ransom, M.~Roberts and J.~Hessels,
  ``Shapiro Delay Measurement of A Two Solar Mass Neutron Star,''
  Nature {\bf 467}, 1081 (2010).

\bibitem{Antoniadis:2013pzd}
  J.~Antoniadis, P.~C.~C.~Freire, N.~Wex, T.~M.~Tauris, R.~S.~Lynch, M.~H.~van Kerkwijk, M.~Kramer and C.~Bassa {\it et al.},
  ``A Massive Pulsar in a Compact Relativistic Binary,''
  Science {\bf 340}, 6131 (2013).


\bibitem{Klahn:2006iw}
  T.~Kl\"ahn, D.~Blaschke, F.~Sandin, C.~Fuchs, A.~Faessler, H.~Grigorian,
  G.~R\"opke and J.~Tr\"umper,
  ``Modern compact star observations and the quark matter equation of state,''
  Phys.\ Lett.\ B {\bf 654}, 170 (2007).


\bibitem{Orsaria:2012je}
  M.~Orsaria, H.~Rodrigues, F.~Weber and G.~A.~Contrera,
  ``Quark-hybrid matter in the cores of massive neutron stars,''
  Phys.\ Rev.\ D {\bf 87}, 023001 (2013).

\bibitem{Klahn:2013kga}
  T.~Kl\"ahn, D.~B.~Blaschke and R.~Lastowiecki,
  ``Implications of the measurement of pulsars with two solar masses for quark
matter in compact stars and HIC. A NJL model case study,''
  Phys.\ Rev.\ D {\bf 88}, 085001 (2013).

\bibitem{Orsaria:2013hna}
  M.~Orsaria, H.~Rodrigues, F.~Weber and G.~A.~Contrera,
  ``Quark deconfinement in high-mass neutron stars,''
  arXiv:1308.1657 [nucl-th].

\bibitem{Shao:2013toa}
  G.~Y.~Shao, M.~Colonna, M.~Di Toro, Y.~X.~Liu and B.~Liu,
 ``Isoscalar-vector interaction and hybrid quark core in massive neutron
  stars,''  arXiv:1305.1176 [nucl-th].

\bibitem{Blaschke:2007ri}
  D.~B.~Blaschke, D.~Gomez Dumm, A.~G.~Grunfeld, T.~Kl\"ahn and N.~N.~Scoccola,
  ``Hybrid stars within a covariant, nonlocal chiral quark model,''
  Phys.\ Rev.\ C {\bf 75}, 065804 (2007).

\bibitem{Blaschke:2013rma}
 D.~Blaschke, D.~E.~Alvarez Castillo, S.~Benic, G.~Contrera and R.~Lastowiecki,
  ``Nonlocal PNJL models and heavy hybrid stars,''
  PoS {\bf ConfinementX}, 249 (2012).

\bibitem{Blaschke:2013ana}
  D.~Blaschke, D.~E.~Alvarez-Castillo and S.~Benic,
  ``Mass-radius constraints for compact stars and a critical endpoint,''
  PoS {\bf CPOD 2013}, 063 (2013);  [arXiv:1310.3803 [nucl-th]].

\bibitem{Alvarez-Castillo:2013spa}
  D.~E.~Alvarez-Castillo, S.~Benic, D.~Blaschke and R.~Lastowiecki,
  ``Crossover transition to quark matter in heavy hybrid stars,''
  arXiv:1311.5112 [nucl-th].

\bibitem{Contrera:2010kz}
   G.~A.~Contrera, M.~Orsaria and N.~N.~Scoccola,
   ``Nonlocal Polyakov-Nambu-Jona-Lasinio model with wave function
   renormalization at finite temperature and chemical potential'',
   Phys.\ Rev.\  D {\bf 82}, 054026 (2010).

\bibitem{Dumm:2005}
  D.~Gomez Dumm and N.~N.~Scoccola,
  ``Characteristics of the chiral phase transition in nonlocal quark models,''
  Phys.\ Rev.\ C {\bf 72}, 014909 (2005).

\bibitem{Noguera:2008prd78}
 S.~Noguera and N.~N.~Scoccola,
 ``Nonlocal chiral quark models with wavefunction renormalization:
 sigma properties and pion-pion scattering parameters,''
 Phys.\ Rev.\  D {\bf 78}, 114002 (2008).

\bibitem{GomezDumm:2008sk}
  D.~Gomez Dumm, D.~B.~Blaschke, A.~G.~Grunfeld and N.~N.~Scoccola,
  ``Color neutrality effects in the phase diagram of the PNJL model,''
  Phys.\ Rev.\  D {\bf 78}, 114021 (2008).

\bibitem{Parappilly:2005ei}
  M.~B.~Parappilly, P.~O.~Bowman, U.~M.~Heller, D.~B.~Leinweber,
  A.~G.~Williams and J.~B.~Zhang,
  ``Scaling behavior of quark propagator in full QCD,''
  Phys.\ Rev.\ D {\bf 73}, 054504 (2006).

\bibitem{Kamleh:2007ud}
  W.~Kamleh, P.~O.~Bowman, D.~B.~Leinweber, A.~G.~Williams and J.~Zhang,
  ``Unquenching effects in the quark and gluon propagator,''
  Phys.\ Rev.\ D {\bf 76}, 094501 (2007).

\bibitem{Kaczmarek:2011zz}
  O.~Kaczmarek, F.~Karsch, E.~Laermann, C.~Miao, S.~Mukherjee, P.~Petreczky,
C.~Schmidt, W.~Soeldner and W. Unger,
  ``Phase boundary for the chiral transition in (2+1)-flavor QCD at small
  values of the chemical potential,''
   Phys.\ Rev.\ D {\bf 83}, 014504 (2011).

\bibitem{Dexheimer:2009va}
  V.~A.~Dexheimer and S.~Schramm,
  ``A Novel Approach to Model Hybrid Stars,''
  Phys.\ Rev.\ C {\bf 81}, 045201 (2010).

\bibitem{Schaefer:2007pw}
  B.~-J.~Schaefer, J.~M.~Pawlowski and J.~Wambach,
  ``The Phase Structure of the Polyakov--Quark-Meson Model,''
  Phys.\ Rev.\ D {\bf 76}, 074023 (2007).

\bibitem{Pagura:2012}
V.~Pagura, D.~G{\'o}mez Dumm and N.~N.~Scoccola,
``Deconfinement and chiral restoration in non-local PNJL models at
zero and imaginary chemical potential'',
Phys. Lett. B {\bf 707}, 76 (2012).

\bibitem{Horvatic:2010md}
  D.~Horvatic, D.~Blaschke, D.~Klabucar, O.~Kaczmarek,
  ``Width of the QCD transition in a Polyakov-loop DSE model,''
  Phys.\ Rev.\ D {\bf 84}, 016005 (2011).

\bibitem{Sakai:2010rp}
  Y.~Sakai, T.~Sasaki, H.~Kouno and M.~Yahiro,
 ``Entanglement between deconfinement transition and chiral symmetry
 restoration,''
  Phys.\ Rev.\ D {\bf 82}, 076003 (2010).


\bibitem{Dutra:2013lya}
  M.~Dutra, O.~Lourenço, A.~Delfino, T.~Frederico and M.~Malheiro,
``Polyakov-Nambu-Jona-Lasinio phase diagrams and quarkyonic phase from order parameters,''
  Phys.\ Rev.\ D {\bf 88}, 114013 (2013)

\bibitem{Blaschke:2010ka}
  D.~B.~Blaschke, F.~Sandin, V.~V.~Skokov and S.~Typel,
  ``Accessibility of Color Superconducting Quark Matter Phases in
  Heavy-ion Collisions,''
  Acta Phys.\ Polon.\ Supp.\  {\bf 3}, 741 (2010).

\bibitem{Shao:2011fk}
  G.~Y.~Shao, M.~Di Toro, V.~Greco, M.~Colonna, S.~Plumari, B.~Liu and
  Y.~X.~Liu,
  ``Phase diagrams in the Hadron-PNJL model,''
  Phys.\ Rev.\ D {\bf 84}, 034028 (2011).


\bibitem{Sasaki:2010jz}
  T.~Sasaki, Y.~Sakai, H.~Kouno and M.~Yahiro,
  ``QCD phase diagram at finite baryon and isospin chemical potentials,''
  Phys.\ Rev.\ D {\bf 82}, 116004 (2010).

\bibitem{Hell:2012da}
  T.~Hell, K.~Kashiwa and W.~Weise,
  ``Impact of vector-current interactions on the QCD phase diagram,''
  J. Mod. Phys. {\bf 4}, 644 (2013).

\bibitem{Sasaki:2012bi}
C.~Sasaki and K.~Redlich,
 ``An Effective gluon potential and hybrid approach to Yang-Mills
 thermodynamics'',
Phys.\ Rev.\ D {\bf 86}, 014007 (2012).

\bibitem{Ruggieri:2012ny}
M.~Ruggieri, P.~Alba, P.~Castorina, S.~Plumari, C.~Ratti and V.~Greco,
``Polyakov Loop and Gluon Quasiparticles in Yang-Mills Thermodynamics'',
Phys.\ Rev.\ D {\bf 86}, 054007 (2012).

\bibitem{Fukushima:2012qa}
K.~Fukushima and K.~Kashiwa,
 ``Polyakov loop and QCD thermodynamics from the gluon and ghost propagators'',
Phys.Lett. B {\bf 723}, 360 (2013).

\bibitem{Radzhabov:2010dd}
  A.~E.~Radzhabov, D.~Blaschke, M.~Buballa and M.~K.~Volkov,
  ``Nonlocal PNJL model beyond mean field and the QCD phase transition,''
  Phys.\ Rev.\ D {\bf 83}, 116004 (2011).

\bibitem{Blaschke:2011ry}
  D.~B.~Blaschke, J.~Berdermann, J.~Cleymans and K.~Redlich,
  ``Chiral condensate and chemical freeze-out,''
  Phys.\ Part.\ Nucl.\ Lett.\  {\bf 8}, 811 (2011);
  Few Body Systems {\bf 53}, 99 (2012).

\bibitem{Turko:2011gw}
  L.~Turko, D.~Blaschke, D.~Prorok and J.~Berdermann,
  ``Mott-Hagedorn Resonance Gas and Lattice QCD Results,''
  Acta Phys.\ Polon.\ Supp.\  {\bf 5}, 485 (2012).

\bibitem{Turko:2013taa}
  L.~Turko, D.~Blaschke, D.~Prorok and J.~Berdermann,
  ``An effective model of QCD thermodynamics,''
  J.\ Phys.\ Conf.\ Ser.\  {\bf 455}, 012056 (2013).

\bibitem{Wergieluk:2012gd}
  A.~Wergieluk, D.~Blaschke, Y.~.L.~Kalinovsky and A.~Friesen,
  ``Pion dissociation and Levinson's theorem in hot PNJL quark matter,''
  Phys. Part. Nucl. Lett. {\bf 10}, 660 (2013).

\bibitem{Dubinin:2013yga}
  A.~Dubinin, D.~Blaschke and Y.~. L.~Kalinovsky,
  ``Pion and sigma meson dissociation in a modified NJL model at finite
  temperature,''
  arXiv:1312.0559 [hep-ph].

\bibitem{Benic:2013tga}
  S.~Benic and D.~Blaschke,
  ``Finite temperature Mott transition in a nonlocal PNJL model,''
  Acta Phys.\ Polon.\ Supp.\  {\bf 6}, 947 (2013).

\bibitem{Benic:2013eqa}
  S.~Benic, D.~Blaschke, G.~A.~Contrera and D.~Horvatic,
  ``Medium induced Lorentz symmetry breaking effects in nonlocal PNJL models,''
  Phys.\ Rev.\ D (to appear); 
  [arXiv:1306.0588 [hep-ph]].
\end{thebibliography}
\end{document}